\documentclass[aps,preprint,nofootinbib]{revtex4}%
\usepackage{amsfonts}
\usepackage{amsmath}
\usepackage{amssymb}
\usepackage{graphicx}%
\providecommand{\U}[1]{\protect\rule{.1in}{.1in}}

\begin{document}
\preprint{ }
\title[Short title for running header]{Field-parametrization dependence of Dirac's method for constrained
Hamiltonians with first-class constraints: failure or triumph? Non-covariant models}
\author{N. Kiriushcheva}
\email{nkiriush@uwo.ca}
\author{P. G. Komorowski}
\email{pkomoro@uwo.ca}
\author{S. V. Kuzmin}
\email{skuzmin@uwo.ca}
\affiliation{The Department of Applied Mathematics, The University of Western Ontario,
London, Ontario, N6A 5B7, Canada}
\pacs{PACS number}

\begin{abstract}
We argue that the field-parametrization dependence of Dirac's procedure, for
Hamiltonians with first-class constraints not only preserves covariance in
covariant theories, but in non-covariant gauge theories it allows one to find
the natural field parametrization in which the Hamiltonian formulation
automatically leads to the simplest gauge symmetry. 

\end{abstract}
\volumeyear{year}
\volumenumber{number}
\issuenumber{number}
\eid{identifier}
\date{\today}
\received{}

\maketitle


\section{Introduction\ }

In physics, one cannot overstate the importance of gauge invariance; all
theories of fundamental interactions have this property. The description of
gauge symmetries in the Lagrangian and Hamiltonian formulations has attracted
considerable interest and has a long history. For covariant theories, natural
variables are those that are true tensors with respect to Lorentz or general
coordinate transformations. If such variables are used, then the theory is
manifestly covariant. In addition, the Hamiltonian analysis and the subsequent
restoration of the gauge transformations from the first-class constraints
(Dirac's conjecture \cite{Diracbook}) leads to results equivalent to those
found using the Lagrangian approach (e.g. the Maxwell and Yang-Mills theories,
and General Relativity (GR) with the natural variable: i.e. metric tensor).
Non-covariant changes of field parametrisation produce different results (e.g.
gauge symmetries, algebra of constrains, etc.), and the most prominent example
is the ADM parametrisation, which was analysed in \cite{PSS, Dirac, Myths}.
The field-parametrisation dependence of the Dirac procedure for systems with
first-class constraints (i.e. gauge invariant systems) allows one to single
out the one particular parametrisation that is consistent with covariance, and
thus find the Lagrangian formulation for which the gauge invariance is related
to Noether's differential identities (DIs) -- a linear combination of
Euler-Lagrange derivatives (ELDs). Such DIs and the corresponding gauge
transformations are covariant.

Working with the Lagrangians of either non-covariant systems or covariant
systems without the restrictions imposed by covariance, one may take various
linear combinations of the known DIs to construct many more DIs that describe
different gauge transformations, as well as any field parametrisation
(invertible change of variables) can be used to rewrite the DIs. Therefore,
many gauge symmetries can be easily constructed at the Lagrangian level, in
all conceivable parametrisations. According to Noether's theorem
\cite{Noether, Noether-eng}, an important characteristic of a gauge theory is
its maximum number of independent DIs, and if the value of this maximum number
is preserved many different combinations of the DIs may be used. At the
Lagrangian level, all of the different gauge symmetries, and all of the
different field parametrisations are independent -- each symmetry can be
written in any parametrisation, and in each parametrisation any symmetry can
be described. Of course these symmetries can have different properties -- to
form or not to form a group. In the simplest case (such as the Maxwell theory)
a commutator of two consecutive gauge transformations is zero; in other cases
it may be characterised by a field-independent structure constant or a
field-dependent structure function. The field-parametrisation dependence of
Hamiltonian formulations of gauge theories causes there to be a different
relationship among the parametrisations and symmetries. All parametrisations
and symmetries obtained in the Lagrangian approach can also be described using
Hamiltonian methods; but unlike the Lagrangian approach, the symmetries become
uniquely related to a chosen field parametrisation. Is this parametrisation
dependence a failure or a triumph of Dirac's method? To answer this question,
this dependence should be analysed and understood.

If all possible symmetries (and gauge transformations) can be classified on
the basis of some criteria (e.g. do they or do they not form a group? The
simplicity of commutators, etc.)\footnote{Merely calling one symmetry
`correct' and another `incorrect' (as for example, in \cite{ShestakovaCQG})
should not be used as a criterion \cite{KKK-2}.}, then the Hamiltonian methods
allow one to find the corresponding parametrisation through the
parametrisation dependence of the Hamiltonian procedure for constrained
systems.\ Often this property is considered a weakness of the Hamiltonian
approach for constrained systems, and various attempts to modify Dirac's
procedure or the more revolutionary proposal that one may not consider
\textquotedblleft the Dirac approach as fundamental and
undoubted\textquotedblright\ appear in the literature (e.g.
\cite{ShestakovaCQG}). The main rationalisation, at the root of such radical
suggestions, is that in particular parametrisations, Dirac's method refuses to
produce the \textquotedblleft correct\textquotedblright\ or \textquotedblleft
expected\textquotedblright\ symmetries; but in our view this behaviour is an
important property that, in the case of covariant theories, rules out
non-covariant changes of field variables \cite{KKK-3}.

The goal of our paper is twofold. Firstly, to illustrate the connection of
field parametrisations to the associated gauge symmetries for the Lagrangian
and Hamiltonian formulations. A change of parametrisation is usually discussed
in conjunction with GR where the calculations are very involved, and where
considering different parametrisations would be technically very difficult. By
using simple examples we may better elucidate the results. Secondly, to
demonstrate the utility of the field-parametrisation dependence for the case
of non-covariant models, where the most natural parametrisation for particular
models can be found in an algorithmic way by using Hamiltonian analysis.

In the next Section we consider the Henneaux-Teitelboim-Zanelli (HTZ)
\cite{HTZ} model and demonstrate a general procedure to find a natural
parametrisation when using the Hamiltonian approach. In Section 3 the
Isotropic Cosmological Model \cite{ShestakovaCQG} is considered as another
example of the application of this procedure, and the effect of a change of
field parametrisation at the Lagrangian and Hamiltonian levels is discussed.
The results are summarised in Conclusion. In Appendix \ref{Appendix_A}, the
restoration of gauge invariance from the first-class constraints of the
Hamiltonian formulation of the HTZ model in the original parametrisation is
performed using the Castellani procedure \cite{Castellani}.

\section{The Henneaux-Teitelboim-Zanelli model}

To illustrate the role played by the field-parametrisation dependence of
Dirac's procedure for Hamiltonians with first-class constraints in finding the
natural (simplest) gauge symmetry in the Lagrangian parametrisation, we shall
first consider a simple model that was introduced and discussed in \cite{HTZ}
and in the book \cite{HTbook} (see p. 88). Despite the simplicity of this
model, the application of the Dirac procedure leads to tertiary constraints.
The analog in field theory, in the sense of the appearance of tertiary
constraints, is found in the Hamiltonian analysis \cite{affine-metric} of the
affine-metric formulation of GR \cite{Einstein1925, Einstein-eng}.

The Lagrangian of the simple model of \cite{HTZ,HTbook} is%
\begin{equation}
L=\frac{1}{2}\left[  \left(  \dot{q}_{2}-e^{q_{1}}\right)  ^{2}+\left(
\dot{q}_{3}-q_{2}\right)  ^{2}\right]  . \label{eqnHTZ1}%
\end{equation}
The majority of readers will note that the variable $q_{1}$ enters the
Lagrangian only once, and in such a form that one must wonder if it would not
have been better to redefine $e^{q_{1}}$ as a new simple variable at the
outset, instead of keeping it as a function. Such a field parametrisation
would be more natural for this Lagrangian; and one might doubt that the form
given by (\ref{eqnHTZ1}) is the natural\ choice to make for any analysis, or
if it could lead to any insight in the study of the gauge symmetries of such a
model. By using a procedure that relies upon the field-parametrisation
dependence of Dirac's Hamiltonian formulation, we show that such suspicions
are correct. Further, this procedure is a general one, and it can be applied
equally well to more complicated Lagrangians.

\subsection{The Hamiltonian Formulation of the HTZ Model}

The Hamiltonian formulation of (\ref{eqnHTZ1}) was considered in \cite{HTZ} by
using the Dirac procedure. We elaborate upon this example because we shall
need the details for our subsequent discussion of the model of \cite{HTZ,
HTbook}. First, perform the Legendre transformation,%
\[
H_{T}=p_{1}\dot{q}_{1}+p_{2}\dot{q}_{2}+p_{3}\dot{q}_{3}-L~,
\]
where $p_{i}=\frac{\partial L}{\partial\dot{q}_{i}}$ are momenta conjugate to
$q_{i}$. After expressing velocities (solvable) in terms of momenta, the total
Hamiltonian follows:%
\begin{equation}
H_{T}=p_{1}\dot{q}_{1}+e^{q_{1}}p_{2}+q_{2}p_{3}+\frac{1}{2}\left(
p_{2}\right)  ^{2}+\frac{1}{2}\left(  p_{3}\right)  ^{2}.\label{eqnHTZ3}%
\end{equation}
The time development of the primary constraint,%
\begin{equation}
\phi_{1}\equiv p_{1}~,\label{eqnHTZ4}%
\end{equation}
leads to the secondary constraint,%
\begin{equation}
\dot{\phi}_{1}=\left\{  \phi_{1},H_{T}\right\}  =-e^{q_{1}}p_{2}\equiv\phi
_{2}~;\label{eqnHTZ5}%
\end{equation}
and, in turn, the time development of the secondary yields the tertiary,%
\begin{equation}
\dot{\phi}_{2}=\left\{  \phi_{2},H_{T}\right\}  =e^{q_{1}}p_{3}\equiv\phi
_{3}~.\label{eqnHTZ6}%
\end{equation}
The time development of the tertiary constraint is proportional to itself,%
\begin{equation}
\dot{\phi}_{3}=\left\{  \phi_{3},H_{T}\right\}  =\dot{q}_{1}\phi
_{3}~.\label{eqnHTZ7}%
\end{equation}
Thus no new constraints appear, and closure is attained. The algebra of
constraints is simple and has the following Poisson Brackets (PBs):%
\begin{equation}
\left\{  \phi_{1},\phi_{2}\right\}  =-\phi_{2}~,\text{ \ \ }\left\{  \phi
_{1},\phi_{3}\right\}  =-\phi_{3}~,\text{\ }\left\{  \phi_{2},\phi
_{3}\right\}  =0.\label{eqnHTZ9}%
\end{equation}
Therefore all of the constraints are first-class. According to the Dirac
conjecture \cite{Diracbook}, a knowledge of all first-class constraints is
sufficient for finding the gauge transformations.

Using the method\footnote{According to the authors of \cite{HTZ}
\textquotedblleft our formalism is capable of handling such cases without
difficulties\textquotedblright;\ but the Castellani algorithm
\cite{Castellani} also leads to the same gauge transformations (see Appendix
\ref{Appendix_A} of this paper).} proposed in \cite{HTZ}, the transformations
were restored by constructing a gauge generator that allows one to find the
transformations of phase-space variables. After the elimination of momenta
(i.e. the back-substitution of momenta in terms of velocities) the Lagrangian
transformations were found in \cite{HTZ, HTbook}:%
\begin{equation}
\delta q_{1}=\ddot{\varepsilon}+2\dot{q}_{1}\dot{\varepsilon}+\ddot{q}%
_{1}\varepsilon+\left(  \dot{q}_{1}\right)  ^{2}\varepsilon,\text{
\ \ \ }\delta q_{2}=e^{q_{1}}\left(  \dot{\varepsilon}+\dot{q}_{1}%
\varepsilon\right)  ,\text{ \ \ \ \ \ \ }\delta q_{3}=e^{q_{1}}\varepsilon
\label{eqnHTZ10}%
\end{equation}
(see also \cite{Masud}, where they were obtained by a different method, and
Appendix \ref{Appendix_A}, where these transformations are derived using the
Castellani algorithm).

Equation (\ref{eqnHTZ10}) describes gauge transformations that uniquely follow
from Dirac's Hamiltonian analysis of Lagrangian (\ref{eqnHTZ1}); and it
correctly reproduces the symmetry of the Lagrangian, which in this case can be
easily and directly checked by performing a variation of the Lagrangian under
transformations (\ref{eqnHTZ10}):%
\begin{equation}
\delta L=0. \label{eqnHTZ10a}%
\end{equation}

\subsection{The Lagrangian Formulation of the HTZ Model}

Noether's second theorem \cite{Noether}, which we shall need in our
discussion, can be used also for more complicated models. If the
transformations are known, then we can restore the DI -- a combination of
Euler-Lagrange derivatives that is identically equal to zero
(\textquotedblleft off-shell\textquotedblright, i.e. without imposing
equations of motion, $ELD=0$). For example, a restoration of the DI from a
known transformation was performed by Schwinger \cite{Schwinger} in a
discussion of the Einstein-Cartan action (see also \cite{Trans}); but the
first appearance of such an approach can at least be traced back to the
earlier work of Rosenfeld \cite{Rosenfeld} (see Eqs. (71, 72))\footnote{Thanks
to Salisbury \cite{Preprint}, the paper became available to non-German
readers.}. Consider%
\begin{equation}
\int\left[  E_{\left(  q_{1}\right)  }\delta q_{1}+E_{\left(  q_{2}\right)
}\delta q_{2}+E_{\left(  q_{3}\right)  }\delta q_{3}\right]  dt=\int
I\varepsilon dt,\label{eqnHTZ11}%
\end{equation}
where $I$ is a DI and $E_{\left(  q_{i}\right)  }$ are ELDs for (\ref{eqnHTZ1}%
) (see Eqs. (41-43) of \cite{Masud}):%
\begin{equation}
E_{\left(  q_{1}\right)  }=\frac{\delta L}{\delta q_{1}}=-\dot{q}_{2}e^{q_{1}%
}+e^{2q_{1}},\label{eqnHTZ12}%
\end{equation}%
\begin{equation}
E_{\left(  q_{2}\right)  }=\frac{\delta L}{\delta q_{2}}=-\dot{q}_{3}%
+q_{2}-\ddot{q}_{2}+e^{q_{1}}\dot{q}_{1}~,\label{eqnHTZ13}%
\end{equation}
and%
\begin{equation}
E_{\left(  q_{3}\right)  }=\frac{\delta L}{\delta q_{3}}=-\ddot{q}_{3}+\dot
{q}_{2}~.\label{eqnHTZ14}%
\end{equation}
By substituting transformation (\ref{eqnHTZ10}) into the left hand side of
(\ref{eqnHTZ11}), and singling out the gauge parameter, one obtains the
corresponding DI%
\begin{equation}
I=\ddot{E}_{\left(  q_{1}\right)  }-2\dot{E}_{\left(  q_{1}\right)  }\dot
{q}_{1}-E_{\left(  q_{1}\right)  }\ddot{q}_{1}+\left(  \dot{q}_{1}\right)
^{2}E_{\left(  q_{1}\right)  }-e^{q_{1}}\dot{E}_{\left(  q_{2}\right)
}+e^{q_{1}}E_{\left(  q_{3}\right)  }\equiv0,\label{eqnHTZ15}%
\end{equation}
an identity that can be verified by the substitution of (\ref{eqnHTZ12}%
)-(\ref{eqnHTZ14}). It is a straightforward procedure to handle expressions of
any complexity since terms of different type can be considered separately.
Direct variation, where in general we must combine terms under the total
derivative(s) is more difficult; in more complicated theories, one might not
be able to recognise the combinations of terms that form a total derivative
(but not in the case of (\ref{eqnHTZ10a})).

Let us check the properties of transformations (\ref{eqnHTZ10}) by calculating
the commutator such that%
\begin{equation}
\left[  \delta_{1},\delta_{2}\right]  \left(
\begin{array}
[c]{c}%
q_{1}\\
q_{2}\\
q_{3}%
\end{array}
\right)  =\left(  \delta_{1}\delta_{2}-\delta_{2}\delta_{1}\right)  \left(
\begin{array}
[c]{c}%
q_{1}\\
q_{2}\\
q_{3}%
\end{array}
\right)  =\delta_{\left[  1,2\right]  }\left(
\begin{array}
[c]{c}%
q_{1}\\
q_{2}\\
q_{3}%
\end{array}
\right)  . \label{eqnHTZ16}%
\end{equation}
For all fields $q_{i}$ the calculation gives%
\begin{equation}
\varepsilon_{\left[  1,2\right]  }=\varepsilon_{2}\ddot{\varepsilon}%
_{1}-\varepsilon_{1}\ddot{\varepsilon}_{2}+2\dot{q}_{1}\left(  \varepsilon
_{2}\dot{\varepsilon}_{1}-\varepsilon_{1}\dot{\varepsilon}_{2}\right)  .
\label{eqnHTZ17}%
\end{equation}

A field appears in the definition of parameter $\varepsilon_{\left[
1,2\right]  }$; this might lead to a possible problem in which the
transformations do not form a group. It is already an indication that some
simpler transformations might exist for this Lagrangian, as in the example of
the Maxwell Lagrangian, which is quadratic in fields; Lagrangian
(\ref{eqnHTZ1}) would also be quadratic in fields and their derivatives if
another parametrisation were to be considered.

\subsection{Finding a Simpler Gauge Symmetry}

Is it possible to find another simpler symmetry? And might this simple
symmetry have a commutator of gauge transformations equal to zero? One DI that
uniquely follows from the Hamiltonian analysis of (\ref{eqnHTZ1}) is known,
i.e. (\ref{eqnHTZ15}); and by using this DI we can construct another DI, which
according to the converse of Noether's second theorem \cite{Noether}\ would
give another symmetry with a simpler commutator, either equal to zero or
without a field-dependent structure function. Note that Noether's second
theorem refers to the maximum number of independent DIs. It is obvious that if
one of the ELDs enters the DI in such a form that it has no field-dependent
coefficients, then the corresponding transformations would depend only on the
gauge parameter, and the commutator of such transformations would be zero.
This result is the simplest possible, and for such a rudimentary model as HTZ
with Lagrangian (\ref{eqnHTZ1}) (which is quadratic if non-linearity is not
introduced by some weird choice of parametrisation) it is the result to
expect. This subject is certainly something to explore. The analogy in field
theory is the Maxwell electrodynamics, with a Lagrangian quadratic in
velocities, where the commutator of the gauge transformations is zero (of
course if someone did not introduce a non-covariant change of variables).
Inspecting (\ref{eqnHTZ15}), we can see that two of the three ELDs appear
once, and it is not difficult to eliminate their field-dependent coefficients
by performing a multiplication of DI (\ref{eqnHTZ15}) by $e^{-q_{1}}$,
\begin{equation}
\tilde{I}=e^{-q_{1}}I=e^{-q_{1}}\ddot{E}_{\left(  q_{1}\right)  }-2e^{-q_{1}%
}\dot{q}_{1}\dot{E}_{\left(  q_{1}\right)  }-e^{-q_{1}}\ddot{q}_{1}E_{\left(
q_{1}\right)  }+e^{-q_{1}}\left(  \dot{q}_{1}\right)  ^{2}E_{\left(
q_{1}\right)  }-\dot{E}_{\left(  q_{2}\right)  }+E_{\left(  q_{3}\right)
}\equiv0\label{eqnHTZ21}%
\end{equation}
(even though the DI is modified, it is still a DI). This short cut was based
on the simple analysis of the DI for this particular model. But in general, if
it is unclear how to modify a DI, then one can multiply it by some function of
the fields of the model to find the transformations, which correspond to the
DI; one may then solve for the function under the condition that it makes the
commutator of gauge transformations equal to zero. If it is not possible to
find such a solution, then a function that preserves the Jacobi identity may
be sought (this approach will be used in the next Section).

Two of the ELDs in (\ref{eqnHTZ21}) now have field-independent coefficients.
Let us seek the corresponding transformations by performing the operation
(\ref{eqnHTZ11}) that leads to (\ref{eqnHTZ15}) in inverse order. Note that we
are working with the same parametrisation as before (the fields are
unchanged), but the DI is different, and so are the corresponding
transformations:
\begin{equation}
\tilde{\delta}q_{1}=e^{-q_{1}}\ddot{\eta},\text{ \ }\tilde{\delta}q_{2}%
=\dot{\eta},\text{ \ \ }\tilde{\delta}q_{3}=\eta.\label{eqnHTZ25}%
\end{equation}
In (\ref{eqnHTZ25}) we call the gauge parameter by $\eta$ (which is
field-independent as was $\varepsilon$ before), and the transformation by
$\tilde{\delta}$.

Some terms cancel out, when performing the calculation with (\ref{eqnHTZ21}),
which suggests a simpler way of presenting the DI (of course, some might have
recognised such a simplification at the previous stage, but to do so for a
more complicated model can be very difficult; further, our goal is to
demonstrate an algorithm that automatically produces this simpler
presentation). From transformations (\ref{eqnHTZ25}), a short form of DI
(\ref{eqnHTZ21}) follows --%
\begin{equation}
\tilde{I}=\frac{d^{2}}{dt^{2}}\left(  e^{-q_{1}}E_{\left(  q_{1}\right)
}\right)  -\dot{E}_{\left(  q_{2}\right)  }+E_{\left(  q_{3}\right)  }%
\equiv0.\label{eqnHTZ27}%
\end{equation}
It is obvious that for all of the fields, the commutator of transformations
(\ref{eqnHTZ25}) is zero, i.e.%
\begin{equation}
\left[  \tilde{\delta}_{1},\tilde{\delta}_{2}\right]  \left(
\begin{array}
[c]{c}%
q_{1}\\
q_{2}\\
q_{3}%
\end{array}
\right)  =0.\label{eqnHTZ28}%
\end{equation}

Note that this is \textit{the same parametrisation} as that of the original
Lagrangian, but \textit{the symmetry is now different, i.e.} $\tilde{\delta}$
(compare (\ref{eqnHTZ10}) and (\ref{eqnHTZ25})); further, transformations
(\ref{eqnHTZ10}) are complicated, and the others, (\ref{eqnHTZ25}), are the
simplest possible (zero commutators). We can find more new DIs for this
parametrisation, obtain various transformations, and check their group
properties; but this one, (\ref{eqnHTZ27}), is the simplest.

\subsection{General Procedure of Finding the Simplest Parametrisation}

If one gauge symmetry is already known for the Lagrangian in the
parametrisation considered, then additional symmetries can be constructed. For
example, a symmetry, which might not be of the simplest form, may be obtained
from the Hamiltonian formulation. We can then construct many other symmetries
and find the simplest one. Is it possible to further simplify (\ref{eqnHTZ25})
and (\ref{eqnHTZ27}), or might there be another parametrisation that, in
addition, will lead to this simple symmetry in the Hamiltonian approach, and
so provide the canonical, or the simplest parametrisation for this model? By
its construction, the Hamiltonian method leads to a symmetry of the
Lagrangian, but only one for each particular parametrisation (some freedom
might remain, and we will discuss this later) with a DI that always starts
from the highest-order time derivative of some ELD without a field-dependent
coefficient (e.g. the Maxwell and Yang-Mills theories, metric gravity and DI
(\ref{eqnHTZ15}) for (\ref{eqnHTZ1})). Let us call such a ELD \textit{the
leading ELD}. This is not the case for DI (\ref{eqnHTZ27}) in this
parametrisation: the leading ELD enters (\ref{eqnHTZ21}) with the coefficient
$e^{-q_{1}}$, and that is why the Hamiltonian method produces another symmetry
that corresponds to DI (\ref{eqnHTZ15}) (i.e. $I=\ddot{E}_{\left(
q_{1}\right)  }+...$); but one may look for another parametrisation to obtain
this same, simple symmetry. It is easy to find a new parametrisation for this
model because the leading ELD in (\ref{eqnHTZ27}) and its coefficient share
the same field; therefore, we can just look for a new parametrisation of this
field without having to consider a mixture of all of the variables of the
model. The general procedure of changing a DI under a change of variables was
discussed in Conclusion of \cite{KKK-3}.

We can attempt to change only one variable. Consider some invertible function%
\[
q_{1}=q_{1}\left(  q\right)  ;
\]
the relation among the ELDs for such a change is,%
\[
E_{\left(  q\right)  }=\frac{\delta L\left(  q_{1}=q_{1}\left(  q\right)
,q_{2},q_{3}\right)  }{\delta q}=\frac{\delta L}{\delta q_{1}}\frac{dq_{1}%
}{dq}=E_{\left(  q_{1}\right)  }\frac{dq_{1}}{dq}.
\]
In equation (\ref{eqnHTZ27}), the removal of the field-dependent coefficient
requires:%
\[
\frac{dq_{1}}{dq}=e^{-q_{1}}.
\]
And by solving this ordinary differential equation (ODE), we obtain the
parametrisation%
\begin{equation}
q=e^{q_{1}},\label{eqnHTZ35}%
\end{equation}
which yields the simplest possible Noether DI,
\begin{equation}
\tilde{I}=\ddot{E}_{\left(  q\right)  }-\dot{E}_{\left(  q_{2}\right)
}+E_{\left(  q_{3}\right)  }\equiv0,\label{eqnHTZ36}%
\end{equation}
and the gauge transformations%
\begin{equation}
\tilde{\delta}q=\ddot{\eta},\text{ \ }\tilde{\delta}q_{2}=\dot{\eta},\text{
\ \ }\tilde{\delta}q_{3}=\eta.\label{eqnHTZ38}%
\end{equation}
Note that (\ref{eqnHTZ38}) and (\ref{eqnHTZ25}) represent the same symmetry,
but for different parametrisations; this situation is distinct from having two
different symmetries (i.e. (\ref{eqnHTZ10}) and (\ref{eqnHTZ25})) for the same
parametrisation. Further, DI (\ref{eqnHTZ27}) is the same as DI
(\ref{eqnHTZ36}); thus we have arrived at a natural, or canonical,
parametrisation of the initial Lagrangian -- a parametrisation that is
consistent with the one that might have been the more logical choice at the
outset. In this particular parametrisation, the Lagrangian (\ref{eqnHTZ1}) is%
\begin{equation}
\tilde{L}=\frac{1}{2}\left[  \left(  \dot{q}_{2}-q\right)  ^{2}+\left(
\dot{q}_{3}-q_{2}\right)  ^{2}\right]  .\label{eqnHTZ40}%
\end{equation}

\subsection{The Hamiltonian Analysis in the New, Canonical Parametrisation}

Let us return to the Hamiltonian formulation. If one starts from Lagrangian
(\ref{eqnHTZ40}), then the total Hamiltonian is%
\begin{equation}
\tilde{H}_{T}=p\dot{q}+qp_{2}+q_{2}p_{3}+\frac{1}{2}\left(  p_{2}\right)
^{2}+\frac{1}{2}\left(  p_{3}\right)  ^{2}\label{eqnHTZ41}%
\end{equation}
with the primary constraint%
\[
\tilde{\phi}_{1}\equiv p,
\]
where $p$\ is a momentum conjugate to $q$. The secondary,%
\[
\left\{  \tilde{\phi}_{1},H_{T}\right\}  =-p_{2}\equiv\tilde{\phi}_{2}~,
\]
and tertiary,%
\[
\left\{  \tilde{\phi}_{2},H_{T}\right\}  =p_{3}\equiv\tilde{\phi}_{3}~,
\]
constraints follow from the conservation in time of the primary and secondary
constraints. Since there are no new constraints, as $\left\{  \tilde{\phi}%
_{3},H_{T}\right\}  =0$, the Dirac procedure closes on the tertiary constraint
(as in the case for Lagrangian (\ref{eqnHTZ1})). Unlike (\ref{eqnHTZ9}), all
of the PBs among the constraints are zero, i.e.%
\begin{equation}
\left\{  \tilde{\phi}_{1},\tilde{\phi}_{2}\right\}  =\left\{  \tilde{\phi}%
_{1},\tilde{\phi}_{3}\right\}  =\left\{  \tilde{\phi}_{2},\tilde{\phi}%
_{3}\right\}  =0,\label{eqnHTZ50}%
\end{equation}
since all the constraints are pure momenta.

For the Lagrangians in two distinct parametrisations, (\ref{eqnHTZ1}) and
(\ref{eqnHTZ40}), the Dirac method gives two different symmetries,
(\ref{eqnHTZ10}) and (\ref{eqnHTZ38}). What is the relationship between the
Hamiltonians of the two different parametrisations? For these two
parametrisations, it is not difficult to find the canonical transformations
between their phase-space variables:%
\begin{equation}
q=e^{q_{1}},\text{ \ \ \ }p=e^{-q_{1}}p_{1}~,\label{eqnHTZ51a}%
\end{equation}
which satisfy the PBs%
\[
\left\{  q,p\right\}  _{q_{1},p_{1}}=\left\{  q_{1},p_{1}\right\}  _{q,p}=1.
\]
Only one field and its momentum were changed, and the rest of the variables
remain the same. This change of variables, (\ref{eqnHTZ51a}), is canonical.

For constrained systems, the ordinary canonicity condition \cite{Lanczos} is
necessary, but not sufficient to have equivalent Hamiltonians (i.e. that lead
to the same gauge symmetry); the algebra of constraints must also be
preserved, which is not the case for the two Hamiltonians (\ref{eqnHTZ3}) and
(\ref{eqnHTZ41}) (compare with (\ref{eqnHTZ9}) and (\ref{eqnHTZ50})). This
additional condition for the two Hamiltonians with first-class constraints was
conjectured in \cite{FKK}, based on a comparison of the Hamiltonian
formulations of GR due to Pirani, Schild, Skinner (PSS) \cite{PSS}, and due to
Dirac \cite{Dirac}, both of which lead to diffeomorphism invariance
\cite{KKRV, Myths}.

In \cite{HTZ, HTbook} it was noticed\textit{ }that if a gauge parameter is
redefined as $\eta=\eta\left(  q_{1},\varepsilon\right)  =e^{q_{1}}%
\varepsilon$, then one can present (\ref{eqnHTZ10}) in the short form
(\ref{eqnHTZ27}). This result is consistent with the statement made in
\cite{HTZ}: \textquotedblleft... the redefinition of the gauge parameter
$\varepsilon e^{q_{1}}\rightarrow\eta$ simplify the form of the gauge
transformations ...\textquotedblright. If this were only a matter of
\textquotedblleft form\textquotedblright\ (i.e. just a shorthand notation),
then it would be the same symmetry; but if we consider (\ref{eqnHTZ10}) and
treat $\eta$ as a field-independent parameter, then it is a different
transformation, as demonstrated by commutators (\ref{eqnHTZ9}) and
(\ref{eqnHTZ50}). Any field-dependent redefinition of parameters is a new
symmetry, which is clear from the DI. Such a redefinition is equivalent to the
multiplication of a DI by some combination of fields; it is a new DI that
produces a new symmetry. For example, consider DI (\ref{eqnHTZ15});
multiplying it by a field-dependent parameter is in effect a construction of a
new DI, as shown in (\ref{eqnHTZ21}). DIs (\ref{eqnHTZ15}) and (\ref{eqnHTZ21}%
) correspond to two different symmetries, one is complicated and the other has
a zero commutator (as in the Maxwell theory). The latter symmetry cannot be
obtained by the Dirac Hamiltonian method -- for this choice of parametrisation
the complicated one follows uniquely.

To illustrate this difference, let us consider one of transformations
(\ref{eqnHTZ38})%
\[
\tilde{\delta}q=\ddot{\eta};
\]
and using change of variables (\ref{eqnHTZ35}), find the transformation of
$q_{1}$%
\begin{equation}
\tilde{\delta}q_{1}=\tilde{\delta}\left(  \ln q\right)  =\frac{1}{q}%
\tilde{\delta}q=e^{-q_{1}}\ddot{\eta},\label{eqnHTZ51}%
\end{equation}
which is a symmetry with a zero commutator (if the parameter $\eta$ is
field-independent). Transformation (\ref{eqnHTZ51}) differs from ones of
(\ref{eqnHTZ10}) that follow from Hamiltonian (\ref{eqnHTZ3}) and give
commutator (\ref{eqnHTZ16}) with parameter (\ref{eqnHTZ17}). 

The \textquotedblleft field dependent redefinition of gauge
parameter\textquotedblright\ is only useful as a formal trick to simplify some
calculations, including the calculation of commutators and double commutators
as in \cite{KKK-1, KKK-3}. If the parameter were to depend on the fields, then
it would also be affected by the change of variables. If we were to calculate
a commutator, the presence of field-dependent parameters would change the
result; therefore, the formal connection (e.g. $\eta=e^{q_{1}}\varepsilon$) is
actually misleading because the corresponding gauge symmetries are in fact different.

Recall the statement at the beginning of Section 4.4 of \cite{HTZ}:
\textquotedblleft The form of gauge transformations is not unique. One can
redefine the gauge parameters ...\textquotedblright.\ If a gauge parameter is
redefined in the same parametrisation, then it will correspond to a different
symmetry, and it is not a difference in \textquotedblleft
form\textquotedblright. The form of a given gauge transformation is not unique
since different parametrisations may be used without causing a change of
symmetry; but if the gauge parameters are redefined it is not a different form
of the same symmetry, it is a different symmetry.

\subsection{Change of Non-Primary Variables}

Let us discuss the conditions under which the form of gauge transformations
may be changed, but the symmetry remains the same (we do not change the gauge
parameters or multiply the DI by some function; to do so would lead to another
symmetry). Any change of parametrisation will make DI (\ref{eqnHTZ36}), which
is the simplest possible, more complicated; but what will be the effect of
such a change on the gauge symmetry? In the Hamiltonian approach, we always
obtain a symmetry with a corresponding DI such that the leading ELD (the one
with the highest order of time derivative) has no field-dependent coefficient.
Let us consider changes of parametrisation that do not affect the variables
that correspond to the leading ELD. We shall call them\textit{ the primary
variables}. They also play an important role in the Hamiltonian analysis
\cite{affine-metric, Darboux}:\textit{ the primary variables} are those which
conjugate momenta enter the primary first-class constraints. The other
variables, which can be eliminated (along with their conjugate momenta) in the
Hamiltonian reduction by solving the second-class constraints, are called
\textit{second-class variables} (or \textit{secondary}, \textit{non-primary}
variables). In the literature the primary variables are more widely referred
to as `gauge' variables, which is a somewhat confusing name since in a gauge
invariant system all variables are involved in a gauge symmetry, i.e. a
Lagrangian is invariant under the gauge transformations of \textit{all} the
variables in configurational space. 

One may wonder if there can still be some freedom to choose the
parametrisation, and to do so in a way that is consistent with both the
Lagrangian and Hamiltonian approaches? We will consider this question in
general, but take the HTZ model as an example. There are two non-leading ELDs
in (\ref{eqnHTZ36}), which give more flexibility; any invertible change of the
corresponding non-leading variables, including their mixture, will not affect
the gauge transformations, thus keeping the Hamiltonians equivalent; the
symmetry remains the same, even though the transformations are of a different form.

Let us define the transformations,%
\begin{equation}
q_{2}=q_{2}(v,w),~~~q_{3}=q_{3}(v,w),\label{eqnHTZ55}%
\end{equation}
and their inverse,%
\[
\nu=v(q_{2},q_{3}),~~~w=w(q_{2},q_{3}),
\]
to which the condition of invertibility applies,%
\begin{equation}
J=\det\left[
\begin{array}
[c]{cc}%
\frac{\partial q_{2}}{\partial v} & \frac{\partial q_{3}}{\partial v}\\
\frac{\partial q_{2}}{\partial w} & \frac{\partial q_{3}}{\partial w}%
\end{array}
\right]  \neq0.\label{eqnHTZ57}%
\end{equation}
Consider the Lagrangian%
\[
L\left(  q,q_{2},q_{3}\right)  =L\left(  q,q_{2}(v,w),q_{3}(v,w)\right)
=L\left(  q,v,w\right)  =L\left(  q,v(q_{2},q_{3}),w(q_{2},q_{3})\right)  ;
\]
performing this change for DI (\ref{eqnHTZ38}), we must express the ELDs of
the initial parametrisation in terms of the ELDs and fields of the new one. We
have%
\begin{equation}
E_{\left(  v\right)  }=\frac{\delta L}{\delta v}=\frac{\delta L}{\delta q_{2}%
}\frac{\partial q_{2}}{\partial v}+\frac{\delta L}{\delta q_{3}}\frac{\partial
q_{3}}{\partial v}=E_{\left(  q_{2}\right)  }\frac{\partial q_{2}}{\partial
v}+E_{\left(  q_{3}\right)  }\frac{\partial q_{3}}{\partial v}%
\label{eqnHTZ58a}%
\end{equation}
and%
\begin{equation}
E_{\left(  w\right)  }=\frac{\delta L}{\delta w}=\frac{\delta L}{\delta q_{2}%
}\frac{\partial q_{2}}{\partial w}+\frac{\delta L}{\delta q_{3}}\frac{\partial
q_{3}}{\partial w}=E_{\left(  q_{2}\right)  }\frac{\partial q_{2}}{\partial
w}+E_{\left(  q_{3}\right)  }\frac{\partial q_{3}}{\partial w}%
;\label{eqnHTZ58b}%
\end{equation}
as in \cite{KKK-3}, we solve (\ref{eqnHTZ58a}) and (\ref{eqnHTZ58b})\ to find
the relation between the Euler-Lagrange derivatives in this new
parametrisation, i.e.%
\[
E_{\left(  q_{2}\right)  }=\frac{E_{\left(  v\right)  }\frac{\partial q_{3}%
}{\partial w}-E_{\left(  w\right)  }\frac{\partial q_{3}}{\partial v}}%
{J},\text{ \ }E_{\left(  q_{3}\right)  }=\frac{\frac{\partial q_{2}}{\partial
v}E_{\left(  w\right)  }-\frac{\partial q_{2}}{\partial w}E_{\left(  v\right)
}}{J}~.
\]
After performing the substitution into (\ref{eqnHTZ36}), we obtain the DI in
terms of the new variables and corresponding ELDs:%
\[
\tilde{I}=\ddot{E}_{\left(  q\right)  }-\dot{E}_{\left(  q_{2}\right)
}+E_{\left(  q_{3}\right)  }=\ddot{E}_{\left(  q\right)  }-\frac{d}{dt}\left(
\frac{E_{\left(  v\right)  }\frac{\partial q_{3}}{\partial w}-E_{\left(
w\right)  }\frac{\partial q_{3}}{\partial v}}{J}\right)  +\frac{\frac{\partial
q_{2}}{\partial v}E_{\left(  w\right)  }-\frac{\partial q_{2}}{\partial
w}E_{\left(  v\right)  }}{J}~;
\]
and from this result, the gauge transformations of the fields follow:%
\begin{equation}
\tilde{\delta}q=\ddot{\eta},\text{ \ }\tilde{\delta}v=\frac{\frac{\partial
q_{3}}{\partial w}}{J}\dot{\eta}+\frac{-\frac{\partial q_{2}}{\partial w}}%
{J}\eta,\text{ \ \ }\tilde{\delta}w=\frac{-\frac{\partial q_{3}}{\partial v}%
}{J}\dot{\eta}+\frac{\frac{\partial q_{2}}{\partial v}}{J}\eta
~.\label{eqnHTZ59}%
\end{equation}
Compared with transformations (\ref{eqnHTZ38}), the new \textit{form} (i.e.
(\ref{eqnHTZ59})) is very different, in general. To show the equivalence of
the two sets of transformations, (\ref{eqnHTZ38}) and (\ref{eqnHTZ59}), let us
find the transformation of $q_{2}$:%
\begin{equation}
\tilde{\delta}q_{2}=\frac{\partial q_{2}}{\partial v}\tilde{\delta}%
v+\frac{\partial q_{2}}{\partial w}\tilde{\delta}w~.\label{eqnHTZ60}%
\end{equation}
After substitution of (\ref{eqnHTZ59}) into (\ref{eqnHTZ60}) and subsequent
simplification, we return to the transformations (\ref{eqnHTZ38}). The same
result can be found for $q_{3}$. Since we did not perform a multiplication of
the DI by a function of fields, the change of variables in (\ref{eqnHTZ55})
does not affect the gauge invariance. This behaviour is expected. If the DI is
modified by changing to another parametrisation in the manner just described,
then the symmetry is unchanged. This is a true difference in the
\textquotedblleft form\textquotedblright\ of the one (the same) symmetry, i.e.
the form of the gauge transformations is changed, but the commutators among
the gauge transformations of the fields remain the same.

The conjecture was made in \cite{FKK} that to have the same symmetry between
different parametrisations at the Hamiltonian level, the canonicity of the
phase-space variables is not sufficient, the whole algebra of the PBs of
constraints should be unchanged.

\subsection{Change of Non-Primary Variables at the Hamiltonian Level}

Let us now investigate the effect of change of variables (\ref{eqnHTZ55}) at
the Hamiltonian level, but without starting from a Lagrangian that is written
in new variables (such calculations are not difficult); instead let us perform
a change of phase-space variables in total Hamiltonian (\ref{eqnHTZ41}).

At the Hamiltonian level, we can find canonical transformations for change
(\ref{eqnHTZ55}) (e.g. see \cite{Lanczos}):%
\[
p_{2}\dot{q}_{2}+p_{3}\dot{q}_{3}=p_{2}\left(  \frac{\partial q_{2}}{\partial
v}\dot{v}+\frac{\partial q_{2}}{\partial w}\dot{w}\right)  +p_{3}\left(
\frac{\partial q_{3}}{\partial v}\dot{v}+\frac{\partial q_{3}}{\partial w}%
\dot{w}\right)  =
\]%
\[
\left(  p_{2}\frac{\partial q_{2}}{\partial v}+p_{3}\frac{\partial q_{3}%
}{\partial v}\right)  \dot{v}+\left(  p_{2}\frac{\partial q_{2}}{\partial
w}+p_{3}\frac{\partial q_{3}}{\partial w}\right)  \dot{w}=\pi_{v}\dot{v}%
+\pi_{w}\dot{w};
\]
and we have%
\[
\pi_{v}=p_{2}\frac{\partial q_{2}}{\partial v}+p_{3}\frac{\partial q_{3}%
}{\partial v}%
\]
and%
\[
\pi_{w}=p_{2}\frac{\partial q_{2}}{\partial w}+p_{3}\frac{\partial q_{3}%
}{\partial w}.
\]
We may express the momenta of the original formulation in terms of the new
variables%
\[
p_{2}=\frac{\pi_{v}\frac{\partial q_{3}}{\partial w}-\pi_{w}\frac{\partial
q_{3}}{\partial v}}{J},\text{ \ }p_{3}=\frac{\frac{\partial q_{2}}{\partial
v}\pi_{w}-\frac{\partial q_{2}}{\partial w}\pi_{v}}{J},
\]
where $J$ is given by (\ref{eqnHTZ57}). We now have a canonical relationship,
and the only non-zero PBs are,%
\begin{equation}
\left\{  q_{2},p_{2}\right\}  _{v,w,\pi_{v},\pi_{w}}=\left\{  q_{3}%
,p_{3}\right\}  _{v,w,\pi_{v},\pi_{w}}=1;\label{eqnHTZ65}%
\end{equation}
the change of variables is canonical, in the ordinary sense, because of the
invertibility of transformations (\ref{eqnHTZ55}). Performing this
transformation in Hamiltonian (\ref{eqnHTZ41}), only the so-called canonical
part is affected by such a change\footnote{The name \textquotedblleft
canonical\textquotedblright\ for the part without primary constraints conveys
the idea that it is possible to perform canonical transformations as in
ordinary unconstrained Hamiltonians, if the primary variables are not involved
in such changes. This is true even in this example where the whole canonical
part is proportional to the constraints (secondary and tertiary).},%
\begin{equation}
H_{T}\left(  p,q,v,\pi_{v},w,\pi_{w}\right)  =p\dot{q}+H_{c}~,\label{eqnHTZ66}%
\end{equation}%
\[
H_{c}=q\frac{\pi_{v}\frac{\partial q_{3}}{\partial w}-\pi_{w}\frac{\partial
q_{3}}{\partial v}}{J}+q_{2}\left(  v,w\right)  \frac{\frac{\partial q_{2}%
}{\partial v}\pi_{w}-\frac{\partial q_{2}}{\partial w}\pi_{v}}{J}+\frac{1}%
{2}\left(  \frac{\pi_{v}\frac{\partial q_{3}}{\partial w}-\pi_{w}%
\frac{\partial q_{3}}{\partial v}}{J}\right)  ^{2}+\frac{1}{2}\left(
\frac{\frac{\partial q_{2}}{\partial v}\pi_{w}-\frac{\partial q_{2}}{\partial
w}\pi_{v}}{J}\right)  ^{2}.
\]
The time development of primary constraint $p$ leads to the secondary constraint,%

\[
\dot{p}=\left\{  p,H_{T}\right\}  =-\frac{\pi_{v}\frac{\partial q_{3}%
}{\partial w}-\pi_{w}\frac{\partial q_{3}}{\partial v}}{J}=\hat{\phi}_{2}~,
\]
and the time derivative of the secondary constraint yields the tertiary one,%

\[
\left\{  \hat{\phi}_{2},H_{T}\right\}  =\frac{\frac{\partial q_{2}}{\partial
v}\pi_{w}-\frac{\partial q_{2}}{\partial w}\pi_{v}}{J}=\hat{\phi}_{3}~.
\]

Because of canonicity relation (\ref{eqnHTZ65}), these constraints have the
same algebra of PBs as (\ref{eqnHTZ50}); and the two Hamiltonians,
(\ref{eqnHTZ41}) and (\ref{eqnHTZ66}), describe the same symmetry. Based on
these constraints, generators can be built, Hamiltonian gauge transformations
found, and the Lagrangian transformations (\ref{eqnHTZ59})\textbf{ }derived.

\subsection{Discussion}

We demonstrated, using the HTZ model, that the field-parametrisation
dependence of the Dirac procedure allows one to find, for a Lagrangian written
in some parametrisation, the canonical parametrisation that leads to the
simplest gauge transformations with trivial group properties. The procedure
described in this Section can be used to treat more complicated models or
theories, and the obvious and natural results obtained for the HTZ model
illustrate the value of such an approach. Canonical symmetry is a unique
property of the Lagrangian, but the choice of parametrisation is not; any
canonical changes of field variables that preserve the algebra of constraints
keeps the symmetry in tact. Changes of variables that involve
primary\ variables, or their mixture with the non-primary variables, will lead
to another symmetry, and thus cannot be performed at the Hamiltonian level
(such changes are non-canonical). We shall illustrate this behaviour with the
example in the next Section.

\section{Isotropic Cosmological Model}

In this Section we shall use the field-parametrisation dependence of Dirac's
method to investigate the canonical (natural) parametrisation of the so-called
isotropic cosmological model (ICM) (its physical meaning, if any, is not the
subject of our discussion). This model is an example of a gauge theory where
the Dirac procedure reaches closure on the secondary first-class constraint,
unlike in the model of the previous Section. The model is described by the
following Lagrangian with two variables $\left(  N,a\right)  $
\cite{ShestakovaCQG},%
\begin{equation}
L_{1}=-\frac{1}{2}\frac{a\dot{a}^{2}}{N}+\frac{1}{2}Na.\label{eqnC1}%
\end{equation}

Its non-linearity\footnote{The description of non-linearity in
\cite{ShestakovaCQG} was made in reference to the equations; i.e. at the
Lagrangian level at least some terms are not quadratic in fields.} cannot be
eliminated by any field redefinition, and the variation of $L_{1}$ under the
gauge transformations produces a total time derivative, rather than being
exactly equal to zero, as is the case of the HTZ model (see (\ref{eqnHTZ10a}%
)). The model represented by Lagrangian (\ref{eqnC1}) attracts considerable
attention, and it is a topic of extensive discussion by Shestakova
\cite{ShestakovaCQG, Shestakova-1, Shestakova-2}; the choice of a
\textquotedblleft correct\textquotedblright\ symmetry for this model
\cite{ShestakovaCQG} (see also \cite{KKK-2}) was used to support the claim
that \textquotedblleft we cannot consider the Dirac approach fundamental and
undoubted\textquotedblright\ \cite{ShestakovaCQG}. A brief discussion of some
aspects of the results in \cite{ShestakovaCQG} was given in the conclusion of
\cite{FKK} and in our comment \cite{KKK-2}, which provides a description of a
method to find the simplest and most natural parametrisation with the simplest
commutator of the gauge transformations, i.e. the \textit{correct} symmetry,
without invoking unjustifiable arguments and approximations that are external
to the model. We shall use this model to show how a parametrisation, which
leads to the simplest symmetry and field variables, can be found in Dirac's approach.

For the two parametrisations,%
\begin{equation}
N=\sqrt{\mu}\label{eqnC2}%
\end{equation}
(see \cite{ShestakovaCQG,KKK-2}) and%
\begin{equation}
N=e^{-\varkappa}\label{eqnC3}%
\end{equation}
(see \cite{KKK-2}), the Dirac procedure yields distinct gauge transformations
that belong to the Lagrangian in all parametrisations. These gauge
transformations are more complicated than the ones for the original
parametrisation (\ref{eqnC1}); and this confirms the conclusion that one may
draw from an analysis of (\ref{eqnC1}) -- the need for such changes is
suspect\textit{ }and unlikely to lead to some simplification\footnote{Such
changes, (\ref{eqnC2}) and (\ref{eqnC3}), convert both terms in the Lagrangian
into non-quadratic expressions (actually they both become non-polynomial),
contrary to formulation (\ref{eqnC1}).}. The same conclusion was drawn for the
model studied in the previous Section. In \cite{ShestakovaCQG}, the author
used the field-parametrisation dependence of the Dirac method to show that it
produces the transformations that were proclaimed `correct'; but because in
another parametrisation the Dirac method did not lead to the same `correct'
transformations, it was declared that \textquotedblleft we cannot consider the
Dirac approach as fundamental and undoubted\textquotedblright%
\ \cite{ShestakovaCQG}.

\subsection{The Hamiltonian and Lagrangian Analyses of ICM}

For our discussion, we need the results of the Hamiltonian analysis of
(\ref{eqnC1}), which we shall briefly describe (see also
\cite{ShestakovaCQG,KKK-2})). Performing the Legendre transformation%
\[
H_{T}^{\left(  1\right)  }=\pi\dot{N}+p\dot{a}-L_{1}~,
\]
and eliminating the velocities, one obtains%
\begin{equation}
H_{T}^{\left(  1\right)  }=\pi\dot{N}-\frac{1}{2}\frac{N}{a}p^{2}-\frac{1}%
{2}Na.\label{eqnC3h}%
\end{equation}
The time development of the primary constraint leads to the secondary
constraint $T^{\left(  1\right)  }$%
\[
\left\{  \pi,H_{T}^{\left(  1\right)  }\right\}  =\frac{1}{2a}p^{2}+\frac
{1}{2}a=T^{\left(  1\right)  };
\]
and because $\dot{T}^{\left(  1\right)  }=\left\{  T^{\left(  1\right)
},H_{T}\right\}  =0$, closure is reached. The total Hamiltonian can then be
presented in compact form,%
\[
H_{T}^{\left(  1\right)  }=\pi\dot{N}-NT^{\left(  1\right)  }.
\]
The algebra of constraints is simple,%
\[
\left\{  \pi,\pi\right\}  =\left\{  T^{\left(  1\right)  },T^{\left(
1\right)  }\right\}  =\left\{  \pi,T^{\left(  1\right)  }\right\}  =0.
\]
Using the Castellani procedure \cite{Castellani}, the gauge generator can be
constructed \cite{ShestakovaCQG},%
\[
G^{\left(  1\right)  }=-T^{\left(  1\right)  }\theta_{1}+\pi\dot{\theta}_{1}~,
\]
which yields the Hamiltonian gauge transformations in phase space. After
substitution of the momenta in terms of velocities, it then leads to the
Lagrangian form of the gauge
transformations\footnote{\label{Footnote_About_Order}The author of
\cite{ShestakovaCQG} used the generator to find the transformations in the
following form $\delta field=\left\{  field,G\right\}  $, which gives a minus
sign in the gauge transformations; to be consistent with the standard form of
a DI, a plus sign is required. The convention of \cite{ShestakovaCQG} does not
affect the results; we can change the sign in DI or incorporate it into a
gauge parameter without making it field-dependent, therefore, the commutator
of two transformations is unchanged.},%
\begin{equation}
\delta_{1}N=\dot{\theta}_{1}\text{ },\text{\ \ \ }\delta_{1}a=\frac{\dot{a}%
}{N}\theta_{1}~.\label{eqnC3t}%
\end{equation}
The commutator of these transformations is of the simplest possible form,%
\begin{equation}
\left[  \delta_{1}^{\prime},\delta_{1}^{\prime\prime}\right]  \left(
\begin{array}
[c]{c}%
N\\
a
\end{array}
\right)  =\left(  \delta_{1}^{\prime}\delta_{1}^{\prime\prime}-\delta
_{1}^{\prime\prime}\delta_{1}^{\prime}\right)  \left(
\begin{array}
[c]{c}%
N\\
a
\end{array}
\right)  =0.\label{eqnC3c}%
\end{equation}

As in the previous Section, the DI that follows from transformations
(\ref{eqnC3t}) is%
\begin{equation}
I^{\left(  1\right)  }=-\dot{E}_{N}^{\left(  1\right)  }+\frac{\dot{a}}%
{N}E_{a}^{\left(  1\right)  }\equiv0,\label{eqnC4}%
\end{equation}
where the ELDs of (\ref{eqnC1}) are:%
\begin{equation}
E_{N}^{\left(  1\right)  }=\frac{\delta L_{1}}{\delta N}=+\frac{1}{2}%
\frac{a\dot{a}^{2}}{N^{2}}+\frac{1}{2}a\label{eqnC4a}%
\end{equation}
and%

\begin{equation}
E_{a}^{\left(  1\right)  }=\frac{\delta L_{1}}{\delta a}=\frac{a\ddot{a}}%
{N}+\frac{1}{2}\frac{\dot{a}^{2}}{N}-\frac{a\dot{a}}{N^{2}}\dot{N}+\frac{1}%
{2}N. \label{eqnC4b}%
\end{equation}
DI (\ref{eqnC4}) can be checked by the direct substitution of (\ref{eqnC4a})
and (\ref{eqnC4b}).

\subsection{Another Choice of Parametrisation}

By using a general procedure, let us show for this model how the
parametrisation, which leads to the simplest symmetry, can be found in the
Hamiltonian approach, and how the natural parametrisation for Lagrangian
(\ref{eqnC1}) will emerge from it. One can start from the parametrisations we
have already considered, (\ref{eqnC2}) and (\ref{eqnC3}); but we prefer to
explore a new parametrisation on the basis of the following observation: in
terms of simplicity, there is but one parametrisation for this
Lagrangian\footnote{Strictly speaking, this change of variables has a
disadvantage -- in the original Lagrangian, one term is quadratic in fields
and another is `non-linear'; with this substitution, both are `non-linear', as
was the case in the two parametrisations (\ref{eqnC2}) and (\ref{eqnC3})
considered before \cite{ShestakovaCQG,KKK-2}.}, i.e.%
\begin{equation}
N=\frac{1}{M}.\label{eqnC4c}%
\end{equation}

Let us use this parametrisation as a starting point, in an illustrative
example of a general procedure for finding a canonical parametrisation.
Changing the variables in (\ref{eqnC1}), we obtain the Lagrangian, $L_{4}$, in
a new parametrisation\footnote{We use the subscript $``4"$ for convenience (in
case if one wishes to compare it with \cite{ShestakovaCQG,KKK-2} where
$L_{2}\left(  \mu,a\right)  $ and $L_{3}\left(  \varkappa,a\right)  $ have
been discussed).},%
\begin{equation}
L_{4}=-\frac{1}{2}a\dot{a}^{2}M+\frac{1}{2}\frac{a}{M}.\label{eqnC5}%
\end{equation}

Were we to investigate this Lagrangian, with no foreknowledge of its
characteristics, we would apply Dirac's algorithm to it, and find its gauge
symmetry by repeating the standard steps that were discussed in detail in
\cite{ShestakovaCQG} for this model. Going to the Hamiltonian, the Legendre
transformation yields,%
\begin{equation}
H_{T}^{\left(  4\right)  }=\pi_{M}\dot{M}+p\dot{a}-L_{4}~;\label{eqnC7}%
\end{equation}
one may then find the primary constraint,%
\begin{equation}
\pi_{M}=\frac{\delta L_{4}}{\delta\dot{M}}=0,\label{eqnC9}%
\end{equation}
and the generalised momentum,%
\begin{equation}
p=\frac{\delta L_{4}}{\delta\dot{a}}=-a\dot{a}M.\label{eqnC10}%
\end{equation}
From (\ref{eqnC10}), the velocity $\dot{a}$ can be expressed in terms of
momentum $p$:%
\begin{equation}
\dot{a}=-\frac{p}{aM}.\label{eqnC12}%
\end{equation}
The substitution of (\ref{eqnC12}) into (\ref{eqnC7}) yields the total
Hamiltonian,%
\begin{equation}
H_{T}^{\left(  4\right)  }=\pi_{M}\dot{M}-\frac{1}{2}\frac{p^{2}}{aM}-\frac
{1}{2}\frac{a}{M}.\label{eqnC14}%
\end{equation}
According to the Dirac procedure, one must consider the time development of
the primary constraints (i.e. consistency condition)%
\[
\dot{\pi}_{M}=\left\{  \pi_{M},H_{T}\right\}  =-\frac{1}{2}\frac{p^{2}}%
{aM^{2}}-\frac{1}{2}\frac{a}{M^{2}}=T^{\left(  4\right)  },
\]
until the closure. Because the time development of $T^{\left(  4\right)  }$
does not produce new constraints, closure is reached, and the total
Hamiltonian can be written in compact form,%
\[
H_{T}^{\left(  4\right)  }=\pi_{M}\dot{M}+MT^{\left(  4\right)  }.
\]
The algebra of constraints is simple, and demonstrates that the constraints
are first-class:%
\begin{equation}
\left\{  \pi_{M},\pi_{M}\right\}  =0,\text{ \ \ }\left\{  T^{\left(  4\right)
},T^{\left(  4\right)  }\right\}  =0,\text{ \ \ }\left\{  \pi_{M},T^{\left(
4\right)  }\right\}  =\frac{2}{M}T^{\left(  4\right)  }~.\label{eqnC18}%
\end{equation}

This knowledge of first-class constraints is enough for us to find the gauge
transformations using the Castellani procedure \cite{Castellani} (which is a
formal implementation of Dirac's conjecture \cite{Diracbook}), and to
construct the generator,%
\begin{equation}
G_{4}=\left(  -\frac{2}{M}\dot{M}\pi_{M}-T^{\left(  4\right)  }\right)
\theta_{4}+\pi_{M}\dot{\theta}_{4}~,\label{eqnC20}%
\end{equation}
in a manner analogous to that used in \cite{ShestakovaCQG} for $L_{1}$ and
$L_{2}$, and for $L_{3}$ in \cite{KKK-2}. The generator allows one to
calculate the gauge transformations of the phase-space variables (see footnote
\ref{Footnote_About_Order} about the convention), and also to find the
Lagrangian transformations, after substituting velocities in terms of momenta
using (\ref{eqnC12}):%
\begin{equation}
\delta_{4}M=\left\{  M,G_{4}\right\}  =-\frac{2}{M}\dot{M}\theta_{4}%
+\dot{\theta}_{4}\label{eqnC21}%
\end{equation}
and%
\begin{equation}
\delta_{4}a=\left\{  a,G_{4}\right\}  =\left\{  a,\frac{1}{2}\frac{p^{2}%
}{aM^{2}}\right\}  =\frac{p}{aM^{2}}=\frac{-a\dot{a}M}{aM^{2}}=-\frac{\dot{a}%
}{M}\theta_{4}~.\label{eqnC22}%
\end{equation}
One can check that this is a symmetry of $L_{4}$ by directly performing a
variation of the Lagrangian:%
\[
\delta_{4}L_{4}=\frac{1}{2}\frac{d}{dt}\left(  a\dot{a}^{2}\theta
-\frac{a\theta}{M^{2}}\right)  .
\]

Alternatively, one may find the Noether DI that corresponds to transformations
(\ref{eqnC21}) and (\ref{eqnC22}),%
\begin{equation}
I^{\left(  4\right)  }=-\dot{E}_{M}^{\left(  4\right)  }-\frac{2}{M}\dot
{M}E_{M}^{\left(  4\right)  }-\frac{\dot{a}}{M}E_{a}^{\left(  4\right)
}\equiv0,\label{eqnC25}%
\end{equation}
as was done in the previous Section, where the ELDs were calculated
for\textbf{ }$L_{4}$. DI (\ref{eqnC25}) may be directly confirmed by the
substitution of the ELDs.

\bigskip

The commutator of transformations (\ref{eqnC21})-(\ref{eqnC22}) is%
\begin{equation}
\left[  \delta_{4}^{\prime},\delta_{4}^{\prime\prime}\right]  \left(
\begin{array}
[c]{c}%
M\\
a
\end{array}
\right)  =\left(  \delta_{4}^{\prime}\delta_{4}^{\prime\prime}-\delta
_{4}^{\prime\prime}\delta_{4}^{\prime}\right)  \left(
\begin{array}
[c]{c}%
M\\
a
\end{array}
\right)  =\delta_{4}^{\prime\prime\prime}\left(
\begin{array}
[c]{c}%
M\\
a
\end{array}
\right)  \label{eqnC26}%
\end{equation}
with a new gauge parameter%
\begin{equation}
\theta_{\left[  \delta_{4}^{\prime},\delta_{4}^{\prime\prime}\right]  }%
=2\frac{1}{M}\left(  \theta_{4}^{\prime\prime}\dot{\theta}_{4}^{\prime}%
-\theta_{4}^{\prime}\dot{\theta}_{4}^{\prime\prime}\right)  .\label{eqnC27}%
\end{equation}
The commutator (\ref{eqnC26}) is non-zero, and the new gauge parameter is
field-dependent; we must calculate the double commutators to determine whether
or not the gauge transformations form an algebra. For such a simple model (as
in the previous Section) it is natural to expect that some simpler symmetries
and corresponding parametrisations are possible, and that the Hamiltonian can
be used to find them. Can we have gauge transformations with a zero
commutator, or at least a gauge parameter of the commutator (see
(\ref{eqnC27})) without field dependence? Unlike the model in the previous
Section (see DI (\ref{eqnHTZ21})), the structure of DI (\ref{eqnC25}) is not
simple enough to allow one to immediately see what kind of manipulations (if
any) can lead to a commutator without field dependence in a new gauge parameter.

\subsection{Finding the Simplest Gauge Symmetry}

Let us seek the simplest symmetry by a general method. Because the leading ELD
and the additional contributions that involve this ELD, depend only on the
same variable, we may obtain a new DI by performing a multiplication of
(\ref{eqnC25}) by some general, unspecified function of $M$, i.e.
\begin{equation}
f\left(  M\right)  I^{\left(  4\right)  }=\tilde{I}^{\left(  4\right)  }.
\label{eqnC30}%
\end{equation}
We have modified the DI, as well as the corresponding gauge symmetry, in the
hope of finding the best parametrisation.

The $\tilde{I}^{\left(  4\right)  }$ is also a DI, and we can calculate the
corresponding transformations (N.B. We keep the parametrisation the same) by
the converse of Noether's second theorem \cite{Noether}. We then try to find
the condition on this unspecified function that leads to a zero commutator for
these transformations. The transformations for (\ref{eqnC30}) are:%
\[
\tilde{\delta}_{4}M=\frac{d}{dt}\left(  \eta f\left(  M\right)  \right)
-\frac{2}{M}\dot{M}\eta f\left(  M\right)
\]
and%
\[
\tilde{\delta}_{4}a=-\frac{\dot{a}}{M}\eta f\left(  M\right)
\]
(these are different transformations, which we shall call $\tilde{\delta}_{4}%
$, and we use a new parameter, $\eta$). For field $a$ we obtain%
\[
\left(  \tilde{\delta}_{4}^{\prime}\tilde{\delta}_{4}^{\prime\prime}%
-\tilde{\delta}_{4}^{\prime\prime}\tilde{\delta}_{4}^{\prime}\right)
a=\frac{\dot{a}}{M}f\left(  M\right)  \left(  \eta^{\prime\prime}\dot{\eta
}^{\prime}-\eta^{\prime}\dot{\eta}^{\prime\prime}\right)  \left(  -\frac
{df}{dM}+2f\left(  M\right)  \frac{1}{M}\right)  .
\]
And for the commutator to equal zero, we must solve the ODE%
\begin{equation}
-\frac{df}{dM}+2f\left(  M\right)  \frac{1}{M}=0.\label{eqnC37}%
\end{equation}
The solutions of (\ref{eqnC37}) are%
\begin{equation}
f=\pm M^{2};\label{eqnC38}%
\end{equation}
and with these values we obtain a new DI and new symmetry in the same
parametrisation (we consider the minus sign)\footnote{Had the plus sign been
used instead, then one would have obtained%
\[
\tilde{I}^{\left(  4\right)  }=-M^{2}\dot{E}_{M}-2M\dot{M}E_{M}-\dot{a}%
ME_{a}=-I^{\left(  4\right)  }\equiv0.
\]
Therefore the gauge transformations would only differ by a sign.}:%
\[
I^{\left(  4\right)  }=M^{2}\dot{E}_{M}+2M\dot{M}E_{M}+\dot{a}ME_{a}\equiv0.
\]
For this DI, the transformations are:%
\begin{equation}
\tilde{\delta}_{4}M=-M^{2}\dot{\eta},\text{ \ \ \ \ }\tilde{\delta}_{4}%
a=\dot{a}M\eta;\label{eqnC40}%
\end{equation}
and this result immediately suggests a simpler form of (\ref{eqnC30}):%
\begin{equation}
\tilde{I}^{\left(  4\right)  }=\frac{d}{dt}\left(  M^{2}E_{M}\right)  +\dot
{a}ME_{a}~.\label{eqnC41}%
\end{equation}

Is it possible to keep this DI and the corresponding gauge symmetry, but using
a different parametrisation convert it to a form that can be obtained in the
Hamiltonian approach? In the leading term of (\ref{eqnC41}), we have a field
and the corresponding ELD under the sign of the derivative; as before (see
previous Section), the search for the needed parametrisation is simplified.
Considering a change of only one variable,%
\[
M=M\left(  \tilde{M}\right)  ,
\]
the following relation among the ELDs can be obtained,%
\[
E_{\tilde{M}}=E_{M}\frac{dM}{d\tilde{M}}.
\]
Compare it with the first term in (\ref{eqnC41}); the condition to have
$\dot{E}_{\tilde{M}}$ only, in the new variables (i.e. without field-dependent
coefficients) is%
\[
\frac{dM}{d\tilde{M}}=-M^{2}.
\]
Solving this ODE, we obtain,%
\[
\tilde{M}=\frac{1}{M};
\]
thus returning to the original parametrisation (\ref{eqnC1}), which is found
to be the best, natural choice for this Lagrangian. We found a parametrisation
in which the commutator of transformations is the simplest; therefore, these
variables are canonical, and they lead to a canonical Hamiltonian formulation
for this model.

\subsection{Change of Non-Primary Variables}

Does any freedom remain in how to represent the simplest symmetry in a way
that is consistent in both the Lagrangian and Hamiltonian approaches? Let us
return to Lagrangian (\ref{eqnC1}), which is written in a canonical
parametrisation. To have a possibility to obtain this symmetry in the
Hamiltonian approach we should not change the variable $N$; to do so would
lead to the appearance of a coefficient for the leading ELD in the DI that
describes this symmetry (see (\ref{eqnC4})), and the Hamiltonian would not
reproduce it. (The Hamiltonian analysis demonstrates that such a
reparametrisation is pathological; the method of finding the canonical
parametrisation was illustrated above.)

The only change that preserves the leading ELD is a change of the remaining,
non-primary variables (as in the previous Section); and in Lagrangian
(\ref{eqnC1}) there is only one: $a$. Let us analyse such a change of variable
(at the Lagrangian level, it keeps the DI the same, preserving the symmetry;
there is no multiplication of the DI by some field-dependent function).

Consider the following general change of the non-primary variable $a$,%
\begin{equation}
a=f\left(  b\right)  ,\label{eqnC51a}%
\end{equation}
with the condition that this transformation is invertible. This change only
affects a non-leading ELD,%
\begin{equation}
E_{b}^{\left(  1\right)  }=\frac{\delta L}{\delta b}=\frac{\delta L}{\delta
a}\frac{\partial a}{\partial b}=E_{a}^{\left(  1\right)  }\frac{\partial
f}{\partial b}.\label{eqnC50}%
\end{equation}
For DI (\ref{eqnC4}), one may re-express the field coefficient in terms of the
new variable to obtain,%
\[
-\dot{E}_{N}^{\left(  1\right)  }+\frac{\dot{a}}{N}E_{\left(  a\right)
}^{\left(  1\right)  }=-\dot{E}_{N}^{\left(  1\right)  }+\frac{1}{N}%
\frac{\partial f}{\partial b}\dot{b}E_{\left(  a\right)  }^{\left(  1\right)
}~,
\]
which leads to a new form of the DI,%
\[
-\dot{E}_{N}^{\left(  1\right)  }+\frac{1}{N}\dot{b}E_{\left(  b\right)
}^{\left(  1\right)  }\equiv0,
\]
where $E_{\left(  b\right)  }^{\left(  1\right)  }$ is taken from
(\ref{eqnC50}).

In this case, even the form of the DI is preserved (compare with the previous
Section where a general change of two non-primary variables was considered),
and the gauge transformations are:%
\[
\delta_{1}N=\dot{\theta}_{1},\text{ \ \ \ \ \ }\delta_{1}b=\frac{\dot{b}}%
{N}\theta_{1}~.
\]
Is it the same symmetry? Using field redefinition (\ref{eqnC51a}) we find%
\[
\delta_{1}a=\frac{\delta f}{\delta b}\delta_{1}b=\frac{\delta f}{\delta
b}\frac{\dot{b}}{N}\theta_{1}=\frac{\dot{a}}{N}\theta_{1}~,
\]
where $\dot{a}=\frac{\partial f}{\partial b}\dot{b}$ was used. So any
invertible change of non-primary variables is permissible; it does not affect
the gauge symmetry and the gauge transformation remains the same. 

We can start from the Lagrangian and obtain the same result by performing the
Hamiltonian analysis; these are simple calculations, which are similar to
those we have already presented in this paper. We shall not repeat them. We
wish to mention that this change, (\ref{eqnC51a}), which is similar to
(\ref{eqnHTZ55}) described in the previous Section, can be made at the level
of the total Hamiltonian by using a canonical change of phase-space variables%
\begin{equation}
p_{a}\dot{a}=p_{a}\frac{\delta f}{\delta b}\dot{b}=p_{b}\dot{b},\text{
\ \ \ \ \ }p_{b}=p_{a}\frac{\delta f}{\delta b};\label{eqnC51}%
\end{equation}
the formulae in (\ref{eqnC51}) preserve the algebra of the first-class
constraints, as was conjectured in \cite{FKK} and illustrated for the slightly
more complicated change of two variables in the previous Section (see
(\ref{eqnHTZ55})).

The change of variable $N$, which does offer any improvement for the few
parametrisations that were considered (see (\ref{eqnC2}), (\ref{eqnC3}), and
(\ref{eqnC4c})), can be repeated in a general form in the Hamiltonian and
Lagrangian approaches, to show that only one parametrisation leads to a zero
commutator. The change of $a$ does not affect the gauge symmetry, and even the
form of the DI is preserved.

For the singular Lagrangian (\ref{eqnC1}), its original parametrisation is the
most natural; it has a simple Noether DI, where the coefficient of the leading
ELD (with the highest order derivative) is not field dependent, and the
commutator of the transformations is the simplest possible (see (\ref{eqnC3c})).

\subsection{Change of Parametrisation -- Mixture of Primary and Non-Primary
Variables}

For completeness, let us consider a change of variables that uses a mixture of
$N$ and $a$. Great attention was paid to such changes in the papers of
Shestakova \cite{Shestakova-1, Shestakova-2, ShestakovaCQG, ShestakovaGandC}
as a prerequisite to the attempt to \textquotedblleft restore a legitimate
status of the ADM parametrisation\textquotedblright\ \cite{ShestakovaCQG}\ as
a representation of the Hamiltonian formulation of metric GR. Let us consider
the effect of such a change in this simple model; but instead of the general
form $N=v\left(  \tilde{N},a\right)  $ of \cite{ShestakovaCQG}, we perform a
simple change,%
\begin{equation}
N=\frac{a}{K}.\label{eqnC60}%
\end{equation}
This change of variables was selected in \cite{ShestakovaCQG} to simplify the
Lagrangian:%
\begin{equation}
L_{5}\left(  K,a\right)  =-\frac{1}{2}K\dot{a}^{2}+\frac{1}{2}\frac{a^{2}}%
{K}.\label{eqnC61}%
\end{equation}

Of course any invertible change of fields in the Lagrangian preserves its
symmetry; this fact can be confirmed by using change of variables
(\ref{eqnC60}) and recalculating the DI that describes the simplest
transformations with a zero commutator. A DI similar to (\ref{eqnC4}) can be
found for Lagrangian (\ref{eqnC61}). To obtain a new form of DI, we need a
relation between the ELDs for both of the variables that follow from%
\begin{equation}
\delta L_{1}=E_{N}^{\left(  1\right)  }\delta N+E_{a}^{\left(  1\right)
}\delta a=E_{K}^{\left(  5\right)  }\delta K+E_{a}^{\left(  5\right)  }\delta
a=\delta L_{5}~.\label{eqnC62}%
\end{equation}
Using (\ref{eqnC60}), we have%
\[
\delta K=-\frac{a}{N^{2}}\delta N+\frac{1}{N}\delta a=-\frac{K^{2}}{a}\delta
N+\frac{K}{a}\delta a,
\]
which after performing a substitution into (\ref{eqnC62}) and making a
comparison of the coefficients of $\delta N$ and $\delta a$ yields%
\[
E_{N}^{\left(  1\right)  }=-\frac{K^{2}}{a}E_{K}^{\left(  5\right)  }%
\]
and%
\[
E_{a}^{\left(  1\right)  }=\frac{K}{a}E_{K}^{\left(  5\right)  }%
+E_{a}^{\left(  5\right)  }.
\]
Substitution of these results into (\ref{eqnC4}) leads to a new form of DI,%
\[
I^{\left(  1\right)  }\left(  K,a\right)  =-\frac{d}{dt}\left(  -E_{K}%
^{\left(  5\right)  }\frac{K^{2}}{a}\right)  +\frac{\dot{a}K}{a}\left(
\frac{K}{a}E_{K}^{\left(  5\right)  }+E_{a}^{\left(  5\right)  }\right)
\equiv0.
\]
Its simplification gives%
\begin{equation}
I^{\left(  1\right)  }\left(  K,a\right)  =\frac{K^{2}}{a}\dot{E}_{K}^{\left(
5\right)  }+\frac{2K}{a}\dot{K}E_{K}^{\left(  5\right)  }+\frac{\dot{a}K}%
{a}E_{a}^{\left(  5\right)  }\equiv0,\label{eqnC63}%
\end{equation}
which corresponds to the following gauge transformations:%
\begin{equation}
\delta_{1}K=-\frac{K^{2}}{a}\dot{\theta}_{1}+\frac{K^{2}}{a^{2}}\dot{a}%
\theta_{1},\text{ \ \ }\delta_{1}a=\frac{\dot{a}K}{a}\theta_{1}%
~.\label{eqnC64}%
\end{equation}

These are the same transformations as $\delta_{1}$ in (\ref{eqnC3t}), but
written for field parametrisation (\ref{eqnC60}) instead. The commutator of
two such transformations is zero, as before (although it is not as obvious as
it was for (\ref{eqnC3t}), where in $\delta_{1}N$ only a gauge parameter is
present). The DI has a field-dependent coefficient in front of the leading
ELD; and in this parametrisation, the Hamiltonian cannot reproduce this
symmetry. 

We now wonder what symmetry would follow from the Dirac procedure. Starting
from Lagrangian (\ref{eqnC61}), let us find the Hamiltonian and restore the
gauge transformations. Performing the Legendre transformation yields%
\[
H_{T}^{\left(  5\right)  }=\tilde{\pi}\dot{K}+p\dot{a}-L_{5}~,
\]
where the primary constraint is%
\[
\tilde{\pi}=0.
\]
By eliminating the velocity in terms of a momentum,
\begin{equation}
p=\frac{\delta L_{5}}{\delta\dot{a}}=-K\dot{a},\text{ \ \ \ \ \ }\dot
{a}=-\frac{p}{K},\label{eqnC75}%
\end{equation}
we obtain%
\begin{equation}
H_{T}=\tilde{\pi}\dot{K}-\frac{1}{2K}\left(  p^{2}+a^{2}\right)
.\label{eqnC76}%
\end{equation}
The time development of the primary constraint is%
\[
\left\{  \tilde{\pi},H_{T}\right\}  =-\frac{1}{2K^{2}}\left(  p^{2}%
+a^{2}\right)  =\chi~,
\]
and the time derivative of the secondary constraint gives%
\begin{equation}
\left\{  \chi,H_{T}\right\}  =\left\{  \chi,\tilde{\pi}\dot{K}\right\}
=-2\frac{\dot{K}}{K}\chi~.\label{eqnC80}%
\end{equation}
Therefore the procedure is closed on the secondary constraint. The Castellani
generator is%
\[
G=G_{\left(  1\right)  }\dot{\theta}_{5}+G_{\left(  0\right)  }\theta_{5}~,
\]
where%
\[
G_{\left(  1\right)  }=\tilde{\pi},
\]
and%
\[
G_{\left(  0\right)  }=-\left\{  G_{\left(  1\right)  },H_{T}\right\}
+\alpha\tilde{\pi}.
\]
The function $\alpha$ can be found by using the condition,%
\[
\left\{  G_{\left(  0\right)  },H_{T}\right\}  =primary~.
\]
Calculations similar to those presented in Appendix \ref{Appendix_A} yield
$\alpha=-2\frac{\dot{K}}{K}$; and the explicit form of the generator is%
\begin{equation}
G=\tilde{\pi}\dot{\theta}_{5}+\left(  -\chi-2\frac{\dot{K}}{K}\tilde{\pi
}\right)  \theta_{5}~,\label{eqnC60a}%
\end{equation}
which results in the following gauge transformations:%
\begin{equation}
\delta K=-\dot{\theta}_{5}+2\frac{\dot{K}}{K}\theta_{5},\text{ \ \ }\delta
a=\frac{\dot{a}}{K}\theta_{5}~.\label{eqnC60b}%
\end{equation}
Note that $\delta a=\left\{  G,a\right\}  $ gives an expression that depends
on a momentum, which can be eliminated using (\ref{eqnC75}) to obtain the
transformations in configurational space for the Lagrangian.

The DI that corresponds to these transformations is%
\[
I^{\left(  5\right)  }=\dot{E}_{K}^{\left(  5\right)  }+2\frac{\dot{K}}%
{K}E_{K}^{\left(  5\right)  }+\frac{\dot{a}}{K}E_{a}^{\left(  5\right)
}\equiv0;
\]
and there is no need to check that this is a DI because it is related to
$I^{\left(  1\right)  }$ from (\ref{eqnC63}) by%
\[
I^{\left(  5\right)  }=\frac{a}{K^{2}}I^{\left(  1\right)  }\left(
K,a\right)  ~.
\]

We note that if the commutator of the transformations was zero for the natural
parametrisation of this model, then in a new parametrisation it is%
\[
\left[  \delta_{5}^{\prime},\delta_{5}^{\prime\prime}\right]  \left(
\begin{array}
[c]{c}%
\tilde{N}\\
a
\end{array}
\right)  =\delta_{5}^{\prime\prime\prime}\left(
\begin{array}
[c]{c}%
\tilde{N}\\
a
\end{array}
\right)  ~,
\]
with the following gauge parameter%
\[
\theta_{\left[  \delta_{5}^{\prime},\delta_{5}^{\prime\prime}\right]  }%
=\frac{2}{K}\left(  \dot{\theta}_{5}^{\prime}\theta_{5}^{\prime\prime}%
-\dot{\theta}_{5}^{\prime\prime}\theta_{5}^{\prime}\right)  ,
\]
which unlike (\ref{eqnC3c}), is now field- dependent.

\subsection{Mixture of Primary and Non-primary Variables at the Hamiltonian
Level}

Finally, let us try to apply a change of variables (\ref{eqnC60}) directly at
the Hamiltonian level. Hamiltonians (\ref{eqnC3h}) and (\ref{eqnC76}) cannot
be canonically related because they describe different symmetries; therefore,
the algebras of constraints are different (see (\ref{eqnC18}) and
(\ref{eqnC80})). But another problem arises even in this simple model if a
change of variables mixes the primary with non-primary variables. To perform a
substitution in the Hamiltonian, we have to find the canonical transformations
in phase space. We consider%
\[
\pi_{N}\dot{N}+p\dot{a}=\pi_{N}\left(  -\frac{a}{K^{2}}\dot{K}+\frac{1}{K}%
\dot{a}\right)  +p\dot{a}=
\]%
\[
-\pi_{N}\frac{a}{K^{2}}\dot{K}+\left(  \pi_{N}\frac{1}{K}+p\right)  \dot{a},
\]
which leads to the following redefinition of momenta:%
\begin{equation}
\pi_{N}=-\pi_{K}\frac{K^{2}}{a}\label{eqnC68}%
\end{equation}
and%
\begin{equation}
p=\tilde{p}-\pi_{N}\frac{1}{K}=\tilde{p}+\pi_{K}\frac{K}{a}.\label{eqnC69}%
\end{equation}

This redefinition automatically preserves the canonical PBs among the two sets
of phase-space variables (as in (\ref{eqnHTZ65}) of the previous Section). But
the substitution of such transformations, (\ref{eqnC68}) and (\ref{eqnC69}),
into the total Hamiltonian introduces momenta, conjugate to the primary
variable, into the canonical part of the Hamiltonian,%
\begin{equation}
H_{T}^{\left(  1\right)  }=-\pi_{K}\dot{K}-\pi_{K}\frac{K}{a}\dot{a}-\frac
{1}{2K}\left(  \tilde{p}+\pi_{K}\frac{K}{a}\right)  ^{2}-\frac{1}{2}%
\frac{a^{2}}{K};\label{eqnC70}%
\end{equation}
further, the time derivatives of the variables, which had been eliminated,
reappear, illustrating the collapse of the Hamiltonian formulation. Note that
similar problems occurred in more complicated cases, e.g. for the change of
the original variables of GR, the components of the metric tensor, to the ADM
variables, which cannot be performed in the total Hamiltonian \cite{Myths}. In
the original, natural parametrisation, the Hamiltonian formulation leads to
diffeomorphism in configurational space \cite{PSS, Dirac}, and for the ADM
parametrisation it produces a different symmetry that does not form a group
\cite{KKK-2, KKK-3}. The change of variables due to ADM, is a mixture of the
primary with non-primary variables (see \cite{Myths} and \cite{affine-metric}%
).\textit{ }

For a change of phase-space variables, which involves a mixture of primary and
non-primary variables as in (\ref{eqnC60}), (\ref{eqnC68}), and (\ref{eqnC69}%
), even having canonical PBs does not preserve the equivalence of the two
Hamiltonians, and some nonsensical results are obtained (see (\ref{eqnC70})).
Even these simple models show that such a change of variables does not to
preserve a gauge symmetry. Yet on the basis of such manipulations, some have
concluded that \textquotedblleft clear proof\textquotedblright\ of the
legitimacy of the ADM variables is given\ \cite{ShestakovaCQG}. For covariant
theories (at least those that are covariant before some non-covariant
parametrisations are introduced) the effect of the field-parametrisation
dependence of the Dirac method is left for future discussion.

\section{Conclusion}

The Euler-Lagrange derivatives of gauge invariant Lagrangians are not
independent; they can be combined into linear combinations, i.e. differential
identities, each of which is identically equal to zero (\textquotedblleft
off-shell\textquotedblright). The important characteristic of a gauge
invariant Lagrangian is the maximum number of independent DIs, which is equal
to the number of gauge parameters \cite{Noether}; but because any combination
of DIs can be constructed, the DIs cannot be uniquely specified. It is
possible for a set of DIs, which describes a particular symmetry, to be
written in different field parametrisations. Alternatively, different DIs
might correspond to the gauge transformations, which may or may not have group
properties (e.g. see \cite{KKK-1,KKK-3}); and even with group properties,
commutators of gauge transformations may be of varying complexity\ (e.g.
compare (\ref{eqnHTZ16})-(\ref{eqnHTZ17}) with (\ref{eqnHTZ28}), and
(\ref{eqnC3c}) with (\ref{eqnC26})-(\ref{eqnC27}); see also \cite{KKK-2}).

For covariant theories, in which natural field parametrisations (where fields
are true tensors) are used, the ELDs, DIs, and corresponding gauge
transformations are also automatically true tensors (or tensor densities).
Therefore, all of them are independent of the choice of coordinates when
written in explicitly covariant form (e.g. if a DI is a vector density, as in
metric GR, it is identically zero in all coordinate systems). Any
non-covariant modifications of ELDs, or the use of non-covariant field
parametrisations, destroys such properties; hence, severe restrictions on
coordinate transformations are required to preserve the validity of ELDs, DIs,
and gauge transformations. In covariant theories the natural restriction of
the results to covariant form leads to covariant gauge transformations; when
supplemented by the requirement that the DIs be of the lowest order in the
derivatives of ELDs, the result is unique (although from a covariant DI, an
additional covariant DI with the highest order of derivatives can be
constructed, as was demonstrated in the conclusion of \cite{KKK-3}). The
Hamiltonian formulation, when based on a Lagrangian written in a natural
parametrisation (covariant variables for covariant Lagrangians), produces
covariant gauge transformations for all variables, as one may return to the
gauge transformations in configurational space. Different parametrisations in
the Hamiltonian approach lead to different symmetries, which in complicated
theories (without reference to covariance) can lose the group properties of
the transformations, or possess group properties but with a commutator of
gauge transformations which has field-dependent structure functions. This is
how the Hamiltonian formulation (which is not covariant by construction) can
be used to find the preferable parametrisation of a covariant theory, based on
the simplest gauge transformation properties, upon which parametrisation is
found to be covariant. In fact, the field-parametrisation dependence of the
Dirac method offers a way to preserve covariance when using the Hamiltonian
method, which is innately non-covariant.

For non-covariant models, the guidance provided by covariance on what choice
of variables is natural, is absent; therefore, the field-parametrisation
dependence of the Dirac procedure becomes more important, since it allows one
to find the natural variables for the Lagrangian being considered (i.e.
variables in which the Hamiltonian formulation leads to the gauge
transformations with the simplest algebra). We have used two examples (see
Sections 2 and 3) to illustrate how to find the natural variables. The
procedure can be described as follows: start with a Lagrangian, which is
written in some original parametrisation, and pass to its Hamiltonian
formulation; find the gauge transformations; use Noether's second theorem
\cite{Noether} to find the corresponding DIs; modify these DIs (while still in
the original parametrisation) by multiplying them by an unspecified function
of field variables; and use the converse of Noether's theorem to find new
gauge transformations that correspond to a new DI; then calculate the
commutator; try to specify the function of field variables by imposing the
condition that the commutator be zero; if it is not possible to do so, then
seek a new field-independent structure constant; if this search fails, then
find a new commutator with a structure function that depends on fields, but
ensure that the gauge transformations form a group (the calculation of a
double commutator and Jacobi identity, as in \cite{KKK-1, KKK-3}, might be required).

Modified DIs, that are not of a simple form (e.g. DI (\ref{eqnHTZ21})), and
the corresponding gauge transformations cannot be obtained through the
Hamiltonian approach; but by performing a change of field variables, a
complicated DI can be simplified (i.e. the coefficient of the leading ELD can
be made field-independent). The corresponding parametrisation is then natural
for the model being considered. Further, the new gauge transformations can
also be obtained in the Hamiltonian approach. In the two examples treated in
this paper, we explicitly demonstrated this procedure; but it can be applied
to any model, and thus it becomes a more technically involved consideration
(e.g. in the ADM parametrisation the Hamiltonian approach yields
transformations without group properties; it would be a Herculean Labour to
try to find a change of variables to bring one back to the natural
parametrisation of the Einstein-Hilbert action, i.e. the metric tensor, for
which the Hamiltonian formulation produces gauge transformations with group properties).

The important role of the \textquotedblleft primary\textquotedblright%
\ variables was demonstrated in \cite{affine-metric, Darboux} (at the
Lagrangian level, primary variables are those for which the corresponding ELDs
enter the DIs with the highest order of time derivative, e.g. see
(\ref{eqnHTZ21})). Any change of fields that involves the
primary\ variables\ profoundly affects the properties of the gauge
transformations, unlike changes that only involve the non-primary variables
(in \cite{affine-metric, Darboux} they were called \textquotedblleft
non-primary\textquotedblright, \textquotedblleft secondary\textquotedblright%
\ or \textquotedblleft second class\textquotedblright). Any canonical change
of secondary variables does not affect the group properties of the
commutators, this behaviour is also observed in non-singular theories and
models (where all variables are actually secondary).

The primary variables differ greatly from non-primary ones, and ordinary
canonicity \cite{Lanczos} for the change of variables is a necessary, but not
a sufficient condition to preserve the symmetry in the two Hamiltonian
formulations; the whole algebra of first-class constraints must also be
preserved. The need for such requirements was found when making the comparison
of the two Hamiltonian formulations of metric GR: PSS \cite{PSS}, and Dirac's
\cite{Dirac} -- both lead to diffeomorphism in configurational space
\cite{Myths, KKRV}. Changes that only involve non-primary variables keeps the
properties of the commutators in tact (e.g. (\ref{eqnC51a})). A mixing of
primary and non-primary variables might drastically affect the result of the
Hamiltonian formulation (e.g. see Section 3). There is, however, one special
class of field parametrisation -- a rescaling of primary variables by
functions of non-primary variables, for example, for the models considered in
Sections 2 and 3, it would be $\tilde{q}=qf\left(  q_{1},q_{2}\right)  $ and
$\tilde{N}=Nf\left(  a\right)  $, respectively. 

Consider the natural parametrisation of (\ref{eqnHTZ40}), for which the
commutator of the gauge transformations is zero, such rescaling will not
change the commutator, although the gauge transformations of the new fields
will be different. Therefore, with this additional freedom, the effect of
various parametrisations for the Hamiltonian formulation allows one to find
the natural parametrisation (with the simplest commutator) only up to such a
rescaling. This additional freedom could possibly be related to so-called
counterexamples to the Dirac conjecture that are discussed in the literature.
Some counterexamples just result from incorrect Hamiltonian and Lagrangian
analyses (e.g. \cite{WLW}), while in some papers, Dirac's conjecture is
defended \cite{RRbook, CDD1, CDD2}. Some examples are more complicated, such
as the widely known Cawley Lagrangian \cite{Cawley}, which could just be a
consequence of using an unnatural parametrisation for the proposed model.
There could be parametrisations that not only do not lead to a gauge symmetry
with the simplest commutator, but even make the application of the Dirac
procedure impossible. The solution to this problem is to consider different
parametrisations, including the rescaling of primary variables. But it happens
that all of the counterexamples just mentioned can be resolved by a better
choice of parametrisation; or equally well, that all such counterexamples can
be seen as some models that are broken by a poor choice of parametrisation.
Further, the models with an occurrence of a square of constraints (e.g.
\cite{Cawley}) were excluded by Castellani in his theorem (in which he proved
the Dirac conjecture), i.e. \textquotedblleft\textit{all the FC }[first-class
constraints]\textit{, except those arising as }$\chi^{n}$\textit{..., are part
of gauge generators...\textquotedblright} \cite{Castellani} (p. 364).

The first-class constraints of the Hamiltonian formulation and the DIs in the
Lagrangian formulation are interrelated, also the gauge generators are linear
in constraints and the DIs are linear combinations of ELDs; therefore, the
appearance of a square or higher power of constraints in the Hamiltonian
analysis of a model is a strong indication of the need to change a
parametrisation. The role of field parametrisation in the counterexamples to
the Dirac conjecture is one aspect of our current investigation, and the
results will be reported elsewhere.

\section{Acknowledgment}

We would like to thank A.M. Frolov, D.G.C. McKeon, and A.V. Zvelindovsky for discussions.

\appendix

\section{\label{Appendix_A}The Henneaux-Teitelboim-Zanelli model using the
Castellani procedure}

Let us first outline the steps to be followed in performing the Castellani
procedure \cite{Castellani}. We have already used this procedure to restore
the gauge invariance of various Hamiltonian formulations (e.g. \cite{KKRV,
Myths, affine-metric, 2D, 3D, lin, Darboux}). In \cite{affine-metric} we
describe the peculiarities that arise when the Castellani procedure is applied
to systems with tertiary constraints, which is also important for the system
under consideration -- the HTZ model.

The generator of gauge transformations for the Hamiltonian, with first-class
constraints for a system where tertiary constraints appear, is given by%
\begin{equation}
G\left(  t\right)  =\varepsilon G_{\left(  0\right)  }+\dot{\varepsilon
}G_{\left(  1\right)  }+\ddot{\varepsilon}G_{\left(  2\right)  }%
~,\label{eqnCAS1}%
\end{equation}
where $\varepsilon$ is the gauge parameter, and $\dot{\varepsilon}$ and
$\ddot{\varepsilon}$ are its first and second derivatives with respect to time
(as in the HTZ model, $\varepsilon$ and $G_{\left(  i\right)  }$ are functions
of time only). The number of gauge parameters and their tensorial dimension
(for covariant theories) are uniquely defined by the number of primary
first-class constraints, so for the formulation considered (HTZ), there is one
gauge parameter, $\varepsilon$. The functions $G_{\left(  i\right)  }$ can be
found through the following iterative procedure (see Eq. (16b)
\cite{Castellani}, and for more details of its application to field theory see
also Sections 5 and 6 of \cite{Castellani}):%
\begin{equation}
G_{\left(  2\right)  }=\phi_{1}~,\label{eqnCAS2}%
\end{equation}%
\begin{equation}
G_{\left(  1\right)  }+\left\{  G_{\left(  2\right)  },H_{T}\right\}
=\alpha\phi_{1}~,\label{eqnCAS3}%
\end{equation}%
\begin{equation}
G_{\left(  0\right)  }+\left\{  G_{\left(  1\right)  },H_{T}\right\}
=\beta\phi_{1}~,\label{eqnCAS4}%
\end{equation}%
\begin{equation}
\left\{  G_{\left(  0\right)  },H_{T}\right\}  =primary~.\label{eqnCAS5}%
\end{equation}
(Here, $\alpha$ and $\beta$ are functions of phase-space variables.) Note that
only primary constraints explicitly enter equations (\ref{eqnCAS2}%
)-(\ref{eqnCAS5}). The function $G_{\left(  2\right)  }$ is uniquely defined
as a primary constraint $\phi_{1}$, while the functions $G_{\left(  1\right)
}$ and $G_{\left(  0\right)  }$ are, in general, not just the secondary or
tertiary constraints. When we calculated the tertiary constraint in
(\ref{eqnHTZ6}), we only kept the contribution that was not proportional to
the constraints found at previous stages in the Dirac procedure. In the
Castellani procedure, however, the complete expression for the time
development of a constraint is needed. For example, for $\left\{  \phi
_{2},H_{T}\right\}  $ in $\left\{  G_{\left(  1\right)  },H_{T}\right\}  $ and
$\left\{  G_{\left(  0\right)  },H_{T}\right\}  $, we must substitute%
\begin{equation}
\left\{  \phi_{2},H_{T}\right\}  =\left\{  -e^{q_{1}}p_{2},H_{T}\right\}
=\phi_{2}\dot{q}_{1}+\phi_{3}~.\label{eqnCAS20}%
\end{equation}
Using the PBs among the first-class constraints, and the total Hamiltonian,
which are given by (\ref{eqnHTZ6}), (\ref{eqnHTZ7}), (\ref{eqnCAS20}), and
(\ref{eqnHTZ3}), we can solve (\ref{eqnCAS3}) and (\ref{eqnCAS4}) for
$G_{\left(  1\right)  }$ and $G_{\left(  0\right)  }$, respectively:%
\begin{equation}
G_{\left(  1\right)  }=-\left\{  G_{\left(  2\right)  },H_{T}\right\}
+\alpha\phi_{1}=-\phi_{2}+\alpha\phi_{1}~,\label{eqnCAS6}%
\end{equation}%
\begin{equation}
G_{\left(  0\right)  }=-\left\{  G_{\left(  1\right)  },H_{T}\right\}
+\beta\phi_{1}=-\left\{  -\phi_{2}+\alpha\phi_{1},H_{T}\right\}  +\beta
\phi_{1}=\phi_{2}\dot{q}_{1}+\phi_{3}-\alpha\phi_{2}+\beta\phi_{1}%
~.\label{eqnCAS21}%
\end{equation}
The time development of $G_{\left(  0\right)  }$ from (\ref{eqnCAS5}) allows
us to find unknown functions $\alpha$ and $\beta$:%
\[
\left\{  G_{\left(  0\right)  },H_{T}\right\}  =\left\{  \phi_{3}+\phi_{2}%
\dot{q}_{1}-\alpha\phi_{2}+\beta\phi_{1},H_{T}\right\}
\]%
\begin{equation}
=\phi_{3}\dot{q}_{1}+\left(  \phi_{2}\dot{q}_{1}+\phi_{3}\right)  \dot{q}%
_{1}-\phi_{2}\ddot{q}_{1}-\alpha\left(  \phi_{2}\dot{q}_{1}+\phi_{3}\right)
+\beta\phi_{2}=primary~.\label{eqnCAS22}%
\end{equation}
Collecting terms with constraints $\phi_{3}$ and $\phi_{2}$ gives us equations
for $\alpha$ and $\beta$:%
\begin{equation}
2\dot{q}_{1}-\alpha=0~,\label{eqnCAS23}%
\end{equation}%
\begin{equation}
\left(  \dot{q}_{1}\right)  ^{2}-\ddot{q}_{1}-\alpha\dot{q}_{1}+\beta
=0~.\label{eqnCAS24}%
\end{equation}
Solving (\ref{eqnCAS23})-(\ref{eqnCAS24}) yields:%
\begin{equation}
\alpha=2\dot{q}_{1}~,~~\beta=\left(  \dot{q}_{1}\right)  ^{2}+\ddot{q}%
_{1}~.\label{eqnCAS25}%
\end{equation}
Finally, the gauge generator (\ref{eqnCAS1}), with $\alpha$ and $\beta$ from
(\ref{eqnCAS25}), is%
\begin{equation}
G=\varepsilon\left[  \phi_{3}-\dot{q}_{1}\phi_{2}+\left(  \left(  \dot{q}%
_{1}\right)  ^{2}+\ddot{q}_{1}\right)  \phi_{1}\right]  +\dot{\varepsilon
}\left(  -\phi_{2}+2\dot{q}_{1}\phi_{1}\right)  +\ddot{\varepsilon}\phi
_{1}~.\label{eqnCAS26}%
\end{equation}

To obtain the gauge transformations, we use $\delta q_{i}=\left\{
G,q_{i}\right\}  $ and the explicit expressions for constraints $\phi_{i}$
(\ref{eqnHTZ4})-(\ref{eqnHTZ6}):%
\[
\delta q_{3}=\left\{  G,q_{3}\right\}  =-\frac{\delta G}{\delta p_{3}}%
=-\frac{\delta\left(  \varepsilon\phi_{3}\right)  }{\delta p_{3}}%
=-\varepsilon\frac{\delta\left(  e^{q_{1}}p_{3}\right)  }{\delta p_{3}%
}=-\varepsilon e^{q_{1}}~,
\]%
\[
\delta q_{2}=\left\{  G,q_{2}\right\}  =-\frac{\delta G}{\delta p_{2}}%
=-\frac{\delta\left(  -\varepsilon\dot{q}_{1}\phi_{2}-\dot{\varepsilon}%
\phi_{2}\right)  }{\delta p_{2}}=\left(  \varepsilon\dot{q}_{1}+\dot
{\varepsilon}\right)  \frac{\delta\left(  -e^{q_{1}}p_{2}\right)  }{\delta
p_{2}}=-\left(  \varepsilon\dot{q}_{1}+\dot{\varepsilon}\right)  e^{q_{1}}~,
\]%
\begin{align*}
\delta q_{1} &  =\left\{  G,q_{1}\right\}  =-\frac{\delta G}{\delta p_{1}%
}=-\frac{\delta\left(  \varepsilon\left(  \left(  \dot{q}_{1}\right)
^{2}+\ddot{q}_{1}\right)  \phi_{1}+2\dot{\varepsilon}\dot{q}_{1}\phi_{1}%
+\ddot{\varepsilon}\phi_{1}\right)  }{\delta p_{1}}\\
&  =-\varepsilon\left(  \left(  \dot{q}_{1}\right)  ^{2}+\ddot{q}_{1}\right)
-2\dot{\varepsilon}\dot{q}_{1}-\ddot{\varepsilon}~.
\end{align*}
These results prove that generator (\ref{eqnCAS26}) produces gauge
transformations (\ref{eqnHTZ10}). There is a difference in sign, but the
authors of \cite{HTZ} employed the convention $\delta q_{i}=\left\{
q_{i},G\right\}  $. We conclude that there is no advantage in using the HTZ
approach for finding gauge transformations as the amount of calculation is the
same as that for the Castellani procedure \cite{Castellani}.

\end{document}